\documentclass[sigconf]{acmart}
\renewcommand\footnotetextcopyrightpermission[1]{} 
\usepackage{hyperref}
\usepackage{balance}

\begin{document}
\title{$\left|\{\text{Math, Philosophy, Programming, Writing}\}\right|=1$}
\author{Attila Egri-Nagy}
\affiliation{Mathematics and Natural Sciences\\Akita International University, Japan}
\email{egri-nagy@aiu.ac.jp}
\begin{abstract}
  Philosophical thinking has a side effect:  by aiming to find the essence of a diverse set of  phenomena, it often makes it difficult to see the differences between them. This can be the case with Mathematics, Programming, Writing and Philosophy itself.
  Their unified essence is  having a shared understanding of the world
  helped by off-loading our cognitive efforts to suitable languages.

\end{abstract}

\maketitle
  \begin{quote}
    ``Whenever I try make a philosophical argument, it transforms into
    a mathematical question, and to answer that I end up writing
    code. My fate.'' -- the author's tweet-sized self-introduction.
    \end{quote}

Mathematicians, philosophers, programmers and writers -- is there
anything common in what they do? On the surface, it seems, they all do
same: sitting in front of computers  and typing on keyboards.
Joking aside, here we claim that these are very similar
activities in their inner workings. Rather, they are the essentially the same, if we investigate them at a suitable
level of abstraction.
By suitable we mean general enough to capture all of them, but not too much,
so it still carries some useful information.

What can we see from such a unifying perspective?
The differences are obvious. The demarcation lines are
institutionalized. In formal education one has to choose between these
professions early on.
Therefore, we want to draw attention to the parallels instead.
This may serve as an antidote for any fixed ideas about the strict borderlines between these fields.
The provocative arguments below can be helpful for people who think in terms
of ``I'm good at $X$ but not at $Y$.''
For instance, learning an abstract and thus seemingly useless piece of
mathematics could improve programming skills. Alternatively, one might
try writing essays as well. In reverse, writing computer programs may
help a writer, since working with a very limited language could give
new appreciation of a natural language.
In any case, the statements below are more tentative than definitive, and meant to
be more like challenges than confrontations.

Cognitive metaphors \cite{lakoff1980metaphors} made their way into
mathematics \cite{lakoff2000mathematics} and into the computer science discourse \cite{Videla2017}. These are knowledge transferring devices. One can view the ideas below as a network of metaphors, with the purpose of sharing knowledge between the fields.

\section{Programming is Writing}

It has been shown that programming and writing share
the same mechanism: ``the translation of a high-level idea into low
level sentences or statements''
\cite{2017ProgIsWriting}. Complementing that analysis, here we look
at the similarity from the perspective of the desired effect of written text. Instead of
looking at the process creating software, we examine the end product,
the result of code execution.

\subsection*{What is programming?} In programming we write text, which on a suitable
computational device creates a process whose outcome is some desirable
output, or its dynamics is some required behaviour.

A legendary textbook begins with pointing out the
similarity of sorcery and computer programming \cite{SICP96}. Writing programs is
like casting spells.
It indeed feels like magic when a non-trivial piece of code finally
starts working as it is intended to. The reason of this feeling could
be that initiating the right sequence of events in the computer is by
no means easy. It is like toppling dominoes, every piece has to be at
the right place.

\subsection*{What is writing?} We write text (books, papers, essays, poems, messages etc.), which in the mind of a reader evokes feelings and conjure ideas, or simply just puts a piece of information there. Again, if the pattern of ideas and emotions is complex enough, great skill is needed to realize it in the reader's head.
The order of the presentation can be crucial (e.g.~novels and textbooks). The style, choosing the right one from the available options, is also tricky to master.

  \subsection*{The connection}

  Writing evokes a mental process. Programming initiates a mechanical process.
  Therefore, writing is like programming  but for  a different
  runtime/hardware. Coding for human minds. The expression,
  programming of the human brain, may induce the wrong, mostly
  political connotations.
  And some may recoil from the idea of reading a book as executing a
  program.
  But these reactions close down unbeaten paths for thinking.

First, there is the argument that learning to program is lot easier
than learning to write. Exactly the opposite what people would
think. Everyone learns how to write essays and very few learn how to
code. This is a prime example of mistaking the familiar with the
easy. Programming languages are designed to exclude ambiguity. Natural
languages thrive on ambiguity, meaning often depending on the context. Sentences are side-effect-full: they can recall totally unpredictable associations.

  Literary text is a form of truly unconventional computing. Libraries
  (the printed book ones) are collections of software designed to run
  on human minds. There are of course compatibility issues, as
  usual. It has to be written in a language that the reader can
  understand. It also has to use an up-to-date version of that
  language. Beyond these problems, executing these pieces of
  `software' is remarkably robust. Two people reading the same book
  have a lot to talk about.

Having a firm idea of the purpose of writing is an idealized situation
in both cases. Though some software development methodologies would
like to proceed from a fully defined specification through a clear-cut
implementation to the working product -- it almost never happens that
way. In the implementation phase we continue to learn more about the
problem domain. Taking this to its limits  we can even start with no
specification  but an idea. Exploratory programming can be done in the brainstorming phase of developing software or  in art projects. 

  \section{Mathematics is a programming language}
  There are deeper arguments for this (see propositions as types,
  proofs as programs \cite{PropositionsAsTypes2015}), but here we
  present a direct and practical reasoning. We argue that mathematical
  and computational problem solving share the same mechanisms:
  decomposing problems into subproblems and then integrating them back
  into final solutions.

\subsubsection*{In math, what do we do when we have no idea about a given
  problem?} We doodle on a piece of paper, play with the mathematical
objects appearing in the problem. This is experimenting, collecting data, making observations on how the mathematical objects behave.
In computational problem solving, we play in a REPL, experimenting with
the data structures mentioned in the problem specification, mobilising
knowledge about the relevant functions.

  \subsubsection*{What do we do when proving a conjecture?}
  We write down the statement as if it was already a theorem. Assuming
  we already have some idea how to prove the statement, we can start
  thinking about the structure of the proof. Now comes the wishful
  thinking. If there was a lemma such and such, that would make the
  proof lot easier. So we set out to state a lemma and prove it. The
  process in practice is of course not this idealistic clean, but wishful thinking and problem decompositions are key ingredients.
  Lemmas in mathematical reasoning are like subroutines in programming
  languages.

In the math philosophy classic `Proofs and refutations'
\cite{lakatos1976proofs} the students and teacher are like developers
writing a function, finding lots of test cases showing that the
implementation is not correct yet.
The idea of decomposing conjectures into subconjectures is the same as
functional decomposition.

  \subsubsection*{The re-implementation game}
  When learning a programming language, we often play this: imagine
  that someone
  maliciously removed a useful feature, a beloved function from the
  language.
  Can we recover from the blow? The solution involves implementing the
  removed function again. This forces the learner to truly understand
  the function. It also shows how one concept can depend on another,
  with the dependence often going both ways.

  In mathematics this is the foundation game.
  What can we remove from mathematics with the condition that we remain able to recover the
  whole known mathematical knowledge?
  The official answer is that set theory is enough \cite{stewart2015foundations}, but one can start
  from the theory of types as well \cite{univalent2013homotopy}.

  \subsubsection*{Superficial similarities and differences}
There are some obvious parallels and discrepancies between math and
programming. These may lead to premature conclusions.
For instance, the sigma sum notation is a for-loop. But this brings
only a partial correspondence to imperative programming languages.

  Computer science educators would point out that math is not the same as programming. When students get to the \texttt{x = x + 1} variable assignment, disaster strikes. Trying to balance the `equation' leads to zero and one being the same.
  This happens within mathematics as well: $\infty=\infty+1$. It is
  just we are dealing with a different type of mathematical objects
  and the usual operation is not admissible. For variable assignment,
  we need a computation model that represents memory explicitly
  (e.g.~register machines) instead of methods for dealing with equations.

\subsection*{Cognitive work saving}
  So far we established that doing mathematics is like
  programming. But surely, there must be a clear difference. Code written in programming languages gets executed on machines,
  thus we get some work done for us by a physical process.
Mathematics on the other hand is not executed on computers (with the notable exception of
  theorem provers). Thus, it
  cannot do cognitive work for us, or can it?

  Programmers sometimes say that mathematics is useless for their
  profession. They might mention Calculus as an example of math taught
  in computer science degree but not used in software
  engineering. Let's look at the derivation rules. First a student
  learns how to calculate the derivatives by using explicit limit
  calculations of continuous functions. Doing this in general leads to the derivation
  rules. The rules are easy to apply, purely symbolic manipulations and require no limit
  calculations. Here is a quote about Leibniz's notation.
  \begin{quote}
``The notation
he had developed for the differential and integral calculus, the notation
still used today, made it easy to do complicated calculations with little
thought. It was as though the notation did the work.'' \cite{davis2011universal}
\end{quote}
\emph{How can notation do the work?} It relies on previously proven
results, so by saving work it appears to be doing work without machinery.
The task has been done by someone else, who proved the
theorems before.
Parallel to this we have software libraries, which are collections of functions and data structures
that we can use without writing them again. They are like collections of
useful lemmas.
Also, we call a programming language feature \emph{declarative}, if it
hides a general mechanism for obtaining a solution, therefore we do
not need to specify how to get to it. Derivation rules in Calculus are
declarative in this sense.
Therefore, Mathematics is a programming language that often does not even need a computer.

\subsection*{What is a computer anyway?}
It is a device that stores, retrieves and
transforms information.
Pen and paper fit this definition.
A sheet of paper can store information as we write on it.
We can also retrieve the information just by looking at it again.
It also does information processing, the identity transformation.
Mathematicians scribbling on a napkin during the conference dinner are already computer users.
External symbolic representations and manipulations for the ease of
calculations -- we have been teaching kids programming all along.
If we are fascinated by digital information storage, pulsating circuits or tiny magnetic dots on a platter, then we have to admire books as well. Ink on paper, equally fascinating physical system that is lot more than cellulose. We are just in a different epoch of information storage.

There seems to be no  clear distinction between how Mathematics is done in a traditional way and the use of computers in research.
You could say that it is easier to look at the page and understand the calculation than staring at screens flashing a large number of symbols and digits.
But is it really so?
How about a calculation spreading over ten pages?

\subsection*{Not enough time.}
There is an argument, that doing mathematics should be a natural thing
for us to do, simply because we had not enough time to evolve some new faculty
of the brain for that \cite{devlin2000math}.
Similarly, computer programming is so young that we did not have time
to develop a completely new way of thinking, so it is probably a reuse of our
mathematical skills.

\section{Mathematics is Philosophy done right}
  This statement might offend both sides simultaneously, which is not the intention. However, a general enough definition of Mathematics forces this viewpoint.

\subsection*{What is Mathematics?} This is not an easy question to answer,
  but there is a shortcut. Instead of trying to determine its subject,
  we can define Mathematics as \emph{precise
  thinking}. Precise means that we define the ideas we talk about with
  just the right amount of information, no more, no less. No matter
  what you want to talk about, if you do it precisely, then you end up
  doing mathematics.

  \subsection*{What is Philosophy?}
  Following the above minimalist definition of Mathematics, we are
  forced to leave `precise' out, and say that Philosophy is about
  thinking.
  There are two reasons for the loss of precision. One is accidental
  (and we are responsible for that),
  the other is fundamental (and we cannot do much about it).

  In Philosophy, the debates are often revolve around terms that are
  not clearly defined. Then, honest attempts try to clarify what has
  been said before, leading to more confusion just by the increased
  number of definitions. This is reminiscent of programmers writing
  yet another library for the same purpose.
  Whenever possible, philosophical discussion should start with
  precise definitions, which are by definition mathematical
  concepts. For instance, philosophical enquiries about the nature of
  computation should start with the concept of structure preserving
  maps (algebraic homo- and isomorphisms and their generalizations, \cite{egri2017finite}).
This is not to say that Mathematics has all the answers.  Precise
thinking is a necessary but not always sufficient condition.

  On the other hand, as a fundamental reason for the lack of
  exactness, Philosophy also tries to
  understand concepts that we don't know how to define. That's why
  math has very little to say about ethical questions, and in a
  somewhat circular manner, the nature of mathematics is quite a
  philosophical mystery.

  \section{Programming is Philosophy}

Our goal is to understand the world around us by creating
an abstract model of it.
The way of Philosophy  is to discuss as many ideas as we can conceive
(both normal and wild), and see what makes sense (or simply sounds good).
The programming approach is to explain to a computer how the world (or some part of it) works.
Explaining helps a lot in understanding someone's own ideas.
Since instructing a computer requires very high precision, programming forces us to understand the phenomenon we are trying to model computationally.
So, the philosophical method and programming are different, but the purpose is the same.
It is great progress that the Computer Science -- Philosophy pair has become an established research topic (compared to the author's undergraduate years, when it was considered to be an invalid combination of majors).

There is another way to see that developing software is really about understanding one slice of our world -- be it the physical world, or our social reality, or the relation network of abstract objects.
When modelling goes wrong, the correspondence between the software and
the modelled reality becomes conspicuous. It is enough to mention web
forms, which always seem to contain a required text field for an
unanswerable question.
The data model is always  only map of reality \cite{kent2012data}.

\subsection*{Typing}

Classification is a basic cognitive tool: we naturally divide things
around us into categories by their types.
Consequently, these divisions appear in our models of the world.
Therefore, type systems are defining constituents of programming
languages.

The usefulness of types is never doubted, but it is often hotly
debated whether we should specify types in our programs explicitly or not (static vs.~dynamic typing).
Considering the importance of types in being precise, it may be a bit
surprising that this is an issue at all. What is the core problem in
this debate? A natural language analogy might be useful here. It is
possible to be precise (e.g.~in legal documents), but most of the time
we can get by using our language in less careful ways.
Aiming to be closer to human expression, some programming languages embrace
ambiguity.
Logical precision and flexibility may seem to be to opposing directions
in the design space of programming languages. They are not (e.g.~type
inference, dependent types), but the combination requires a steeper
learning curve.

Strangely, in mathematical publications we mainly use typesetting to
indicate types, which puts the current practice more on the dynamic
typing side. Type checking is done manually by reviewers and readers
of mathematical papers.

\subsection*{Code is an evolving description}

Writing code is learning in a way in which evolution is learning.
The genome describes how to (re)build an organism that is capable of surviving in a given environment.
The better it is adapted to the environment, the higher probability for survival.
So, in a sense, the genome is a description of the environment that is
continuously improved through the evolutionary process \cite{infoevobio}.

Now take a mathematical problem, or a mathematical structure to be investigated.
We start with a simple algorithm that explores the structure.
But it does not work.
Actually, it does, it just takes so much time to finish.
Maybe a few times the age of the universe.
The next step is obvious: find the bottleneck, figure out where time is leaking.
The speed-up comes from recognizing some redundancy in the structure, there are things that needs to be checked only once.
Some parts of the structure that look the same by viewing it from different directions.
In its purest form the redundancy is some symmetry.
Now the algorithm is more efficient, the code contains more information about the structure, as it is exploiting some property.
Run it again, look for the next bottleneck.
Each speed-up exploits some property of the structure. The more we know the faster we calculate. Knowledge is power.

Iterating this process we end up with an efficient algorithm to
compute the given mathematical structure.
The code is morphing into a sophisticated algorithm that is eventually a faithful representation of the original structure. Like a molding process, the code is the complementary shape of the mathematical object, and ultimately they are indistinguishable.

\section{Is there anything else to be unified?}

Whenever we unify different things, we also have to make sure that we do not unify too much.
It would be rather unhelpful to collapse everything into one concept,
so we have to draw a line somewhere.
Of course, such a boundary always ends up as an artificial one. Here, it reflects the scope of
the paper, rather than reality.

\emph{Art?} Partial correspondence can be made between art and programming. Another classic textbook is titled as `The Art of Computer Programming' \cite{TAOCPvol1}. In this case art refers to the aim of perfection, attention to the details of the craft.

In most computer-generated artworks the program is just a
tool. Making the source code part of the piece creates a tangled
phenomenon \cite{2016code}. Unravelling these require quite a sholarly
effort \cite{montfort201310}.

\emph{Music?} Live-coding is now an established form of musical performance.
Is the source code an artifact, or is it only the music that matters?
Music lacks the language of discrete symbols (it has a more powerful,
but less rationally understandable one).

\emph{Spirituality?} How about zen-buddhism? Most modern languages feature koan-style exercises, and refer to meditating over a piece of code.
This appears to be a pop culture reference as the purpose of the meditation is to get rid of any language.

Unifying these with programming is possible, but it may require an
extraordinary personal achievement \cite{lispaikidojazz}.
\section{Conclusion}

Our understanding of the world is coded in natural and artificial
languages. These languages not only store and share information but to
a varying extent they do cognitive work as well. Programming languages
excel in this respect, as they are hooked to physical systems
(computing hardware) to amplify this effect.
The differences in language usages (artificial vs.~natural, declarative vs.~procedural) are not fundamental. One can and should move between them.

We can draw a practical conclusion for education (both institutional and
self-learning) and  repeat old  questions (e.g.~\cite{GEB}) for research (multi- and inter-disciplinary).
\begin{enumerate}
\item Thinking with the help of formal systems - this natural and
  cultural skill should be taught explicitly.
\item How can formal systems do cognitive work? What exactly is computation?
\end{enumerate}

\balance

\begin{acks}
Many thanks to James Noble and an anonymous reviewer, who read the
first draft and gave ample feedback.
\end{acks}

\bibliography{prog}
\bibliographystyle{plain}

\end{document}